# Modelling the thermo-mechanical volume change behaviour of compacted expansive clays

**Anh-Minh Tang and Yu-Jun Cui**

*Researcher and Professor respectively. Ecole des Ponts – ParisTech, UR Navier/CERMES, 6 & 8 av Blaise Pascal, Champs-sur-Marne, 77455 Marne-la-Vallée Cedex 2*

Written date: December 12, 2008

**Corresponding author:**
Prof. Yu-Jun CUI
Ecole Nationale des Ponts et Chaussées, CERMES.
6-8 av. Blaise Pascal, Cité Descartes, Champs-sur-Marne
F-77455 MARNE LA VALLEE
France

Telephone : +33 1 64 15 35 50
Fax : +33 1 64 15 35 62
E-mail : cui@cermes.enpc.fr

**Abstract**: Compacted expansive clays are often considered as a possible buffer material in high-level deep radioactive waste disposals. After the installation of waste canisters, the engineered clay barriers are subjected to thermo-hydro-mechanical actions in the form of water infiltration from the geological barrier, heat dissipation from the radioactive waste canisters, and stresses generated by clay swelling under almost confined conditions. The aim of the present work is to develop a constitutive model that is able to describe the behaviour of compacted expansive clays under these coupled thermo-hydro-mechanical actions. The proposed model is based on two existing models: one for the hydro-mechanical behaviour of compacted expansive clays and another for the thermo-mechanical behaviour of saturated clays. The elaborated model has been validated using the thermo-hydro-mechanical test results on the compacted MX80 bentonite. Comparison between the model prediction and the experimental data show that this model is able to reproduce the main features of volume changes: heating at constant suction and pressure induces either expansion or contraction; the mean yield stress changes with variations of suction or temperature.

# Introduction

Compacted expansive clays are often proposed as engineered barriers and seals in the design of deep geological radioactive waste repositories. In this design, the barriers are made up of highly compacted expansive clays. After the installation of waste canisters, the engineered barriers and seals are subjected at the same time to heating induced by the heat-emitting waste and to water infiltration from the geological barrier. Large-scale tests and numerical modelling works have shown strong coupling between the thermal, hydraulic and mechanical behaviours (Rutqvist *et al.*, 2001; Dixon *et al.*, 2002; Alonso *et al.*, 2005; Hökmark *et al.*, 2007). Various laboratory works have been conducted to study the hydro-mechanical behaviour of compacted expansive soils (Delage *et al.*, 1998; Lloret *et al.*, 2003; Cuisinier & Masrouri, 2004). The main behaviours observed such as the suction effect on the mechanical properties and the soil volume change upon hydro-mechanical loading have been satisfactorily described by various constitutive models (Alonso *et al.*, 1999; Cui *et al.*, 2002; Lloret *et al.*, 2003; Sanchez *et al.*, 2005).

The thermo-mechanical behaviour is often investigated for saturated soils (Delage *et al.*, 2000; Sultan *et al.*, 2002; Cekerevac & Laloui, 2004; Abuel-Naga *et al.*, 2007). These studies showed that heating may induce expansion at large overconsolidation ratios (*OCR*) and contraction at an *OCR* close to one. In addition, heating slightly reduced the mean yield stress. Several constitutive models have been elaborated to describe these experimental observations (Robinet *et al.*, 1996; Cui *et al.*, 2000; Graham *et al.*, 2001; Laloui & Cekerevac, 2003; Abuel-Naga *et al.*, 2007).

Recent experimental works on the thermo-mechanical behaviour of unsaturated expansive soils showed similar behaviour to that of saturated clays (Romero *et al.*, 2005; Lloret & Villar, 2007; Tang *et al.*, 2008). In terms of constitutive modelling, there have been several thermo-mechanical constitutive models developed based on the observations on soils of low plasticity (Saix & Jouanna, 1989; Modaressi & Modaressi, 1995; Devillers *et al.*, 1996). Nevertheless, Gens & Alonso (1992) showed that predicting the mechanical behaviour of expansive clays requires introduction of more complex mechanisms. In terms of numerical modelling, several works have been performed to model the thermo-hydro-mechanical behaviour of compacted expansive clays used as engineered barrier in geological radioactive waste disposals (Thomas & Cleall, 1999; Börgesson *et al.*, 2001; Rutqvist *et al.*, 2001; Sanchez *et al.*, 2005; Alonso *et al.*, 2005; Hökmark *et al.*, 2007). However, the main thermo-mechanical behaviour such as the reduction of mean yield stress upon heating or the contraction when heating at large *OCR* were not taken into account in these studies.

In the present work, a constitutive model is developed to predict the thermo-mechanical behaviour of unsaturated compacted expansive soils. This model is based on the model proposed by Alonso *et al.* (1999) for the hydro-mechanical behaviour of unsaturated expansive soils and on the model proposed by Cui *et al.* (2000) for the thermo-mechanical behaviour of saturated clays. The validation of the model is made based on the experimental results presented by Tang *et al.* (2008) and Tang & Cui (2006) obtained on compacted MX80 clay. Note that Sanchez *et al.* (2007) also introduced the temperature effects on the behaviour of expansive clays following a similar procedure. They modified the double structure expansive model by changing the microstructural stiffness, the LC yield locus and the cohesion changes with suction. The validation of the model was performed by reproducing the dependence of swelling pressure on temperature for two dry densities.

## Model formulation

### Double structure concept

The proposed model is based on the conceptual approach proposed by Gens & Alonso (1992): the soil structure is divided into two levels, a microstructural level and a macrostructural one. The microstructural level corresponds to the clay grains composed of an anisotropic aggregate of clay particles, whereas the macrostructural level corresponds to the larger-scale soil structure (Kröhn, 2003). According to this concept, the void ratio ($e$) can be divided into two parts: micro- and macrostructural void ratios ($e_m$ and $e_M$ respectively), with

$$e = e_m + e_M \qquad [1]$$

### Microstructural volume change

The microstructural pores are assumed to be always saturated. This assumption is supported by the experimental observation made by Saiyouri *et al.* (2000), showing that increasing suction ($s$) reduces the interlayer distance and the interlayer can be considered saturated even at a suction as high as 100 MPa in the case of bentonites. Furthermore, it has been observed that the effects of suction ($s$) and net mean stress ($p$) on microstructural volume changes are similar; the microstructural behaviour is reversible (Gens & Alonso, 1992). Thus, the microstructural volumetric elastic strain is calculated as follows:

$$d\varepsilon_{vm}^e = \frac{de_m}{1+e_m} = \frac{d(p+s)}{K_m} + \alpha_m^T dT \qquad [2]$$

In this formulation, a change in $s$ induces the same volume change as that induced by a change in $p$, i.e., Terzaghi's effective stress concept holds. The microstructural coefficient of thermal expansion, $\alpha_m^T$, is considered to be independent of $p$, $s$ and $T$. The microstructural bulk modulus, $K_m$, is calculated as follows:

$$K_m = \frac{\exp(\alpha_m(p+s))}{\beta_m} \qquad [3]$$

Following this formulation, $K_m$ increases with an increase of $p$ or $s$ and is independent of $T$. This expression was adopted based on the consideration that the mechanical behaviour at microstructural level is controlled by the physico-chemical effects occurring at clay particle level that are basically reversible (Sanchez et al. 2007).

### Macrostructural volume change

The macrostructural elastic volumetric strain is expressed as a function of $p$, $s$ and $T$:

$$d\varepsilon_{vM}^e = \frac{de_M}{1+e_M} = \left(\frac{\kappa}{1+e_M}\right)\frac{dp}{p} + \left(\frac{\kappa_s}{1+e_M}\right)\frac{ds}{s+p_{atm}} + \alpha_M^T dT \qquad [4]$$

where $\kappa$ and $\kappa_s$ are the macrostructural elastic compressibility parameters for changes in net mean stress and in suction respectively, $p_{atm}$ is the atmospheric pressure. In this formulation, the effect of $s$ and $p$ on the elastic volumetric strain is similar to that proposed by Alonso et al. (1990) for soils of low plasticity. The macrostructural coefficient of thermal expansion, $\alpha_M^T$, is also considered to be independent of $p$, $s$ and $T$.

The mechanical loading yield surface (L) is constituted of the LC curve proposed by Alonso *et al.* (1999) in the *s-p* plot and the LY curve proposed by Cui *et al.* (2000) in the *T-p* plot. A schematic view of the surface is presented in Fig. 1. The net mean yield stress $p_0$ is expressed as a function of $s$ and $T$:

$$p_0 = p_c \left(\frac{p_0^*}{p_c}\right)^{\frac{\lambda(0)-\kappa}{\lambda(s)-\kappa}} \exp(-\alpha_0(T-T_0)) \qquad [5]$$

where $p_c$ is a reference stress and $p_0^*$ is the net mean yield stress at zero suction and initial temperature ($T_0$). In this expression, the effect of $T$ on $p_0$ is represented by the coefficient $\alpha_0$. This coefficient is assumed to be independent of $s$ and $p$. The macrostructural compressibility parameter for changes in $p$ for a virgin state at a suction $s$ is calculated as follows:

$$\lambda(s) = \lambda(0)[r + (1-r)\exp(-\beta s)] \qquad [6]$$

where $r$ and $\beta$ are soil constants, $\lambda(0)$ is the macrostructural compressibility parameter for changes in $p$ for a virgin state at zero suction.

The thermal loading yield surface is developed from the TY curve proposed by Cui *et al.* (2000). The yield temperature $T_{CT}$ is calculated as follows:

$$T_{CT} = \left[(T_c - T_0)\exp(-\beta_T \frac{p}{p_0}) + T_0\right]\exp(\alpha_s(s-s_0)) \qquad [7]$$

where $T_c$ is a reference temperature and $s_0$ is the initial suction. In this formulation, $\beta_T$ defines the effect of $p/p_0$ on the yield temperature, which is similar to the effect of the overconsolidation ratio (*OCR*) in the case of saturated soils (Cui et al. 2000): the value of $T_{CT}$ is smaller at lower $p_0/p$ ratio. On the other hand, $\alpha_s$ defines the suction effect: the value of $T_{CT}$ is higher at higher suctions.

Regarding the hydric loading (suction changes), the model considers the two surfaces (Fig. 1) proposed by Alonso *et al.* (1999): SI (Suction Increase) and SD (Suction Decrease). The effect of temperature on these surfaces is not taken into account in the present work. In addition, the two surfaces are assumed to coincide with the Neutral Line (NL); this means that a microstructural deformation will always induce an irreversible macrostructural deformation. With this assumption, it is also as if the SI and SD surfaces are masked by NL. Note that Sanchez et al. (2007) also considered NL solely to separate microstructural swelling from microstructural compression. During hydro-mechanical loadings, the NL moves following the current net mean stress (p) and current suction (s).

The following irreversible volumetric strains are taken into account in the model.
1) When the yield surface L is reached by a mechanical loading:

$$d\varepsilon_{vML}^p = \frac{\lambda(s)-\kappa}{1+e_M}\frac{dp_0}{p_0} \qquad [8]$$

2) Macrostructural plastic strain induced by microstructural strain when SI or SD surfaces are activated:

$$d\varepsilon_{vMSI}^p = f_I d\varepsilon_{vm}^e \qquad [9]$$

$$d\varepsilon_{vMSD}^p = f_D d\varepsilon_{vm}^e \qquad [10]$$

where $f_I$ and $f_D$ are the interaction functions between micro- and macrostructural levels in case of suction increase and suction decrease, respectively.

In the proposed model, the interaction between micro- and macrostructural levels involves only the hydric and mechanical loading as described by Alonso *et al.* (1999). The effect of thermal microstructural strain on the macrostructural strain is omitted due to the lack of experimental evidence. On the other hand, this assumption is also justified by the experimental result showing that thermal strain is negligible when compared to the hydro-mechanical strain (Lloret & Villar, 2007). Therefore, in equations [9] and [10] the temperature term is not introduced.

3) When the thermal yield surface (TY) is reached by a thermal loading (heating):

$$\mathrm{d}\varepsilon^p_{vMT} = \frac{-\alpha^T_M}{1-a}\left[\exp\left(\frac{-\alpha^T_M}{1-a}(T - T^0_{CT})\right) - a\right]\mathrm{d}T \qquad [11]$$

This formulation is similar to that proposed by Cui *et al.* (2000) and the parameters ($\alpha^T_M$ and $a$) are assumed to be independent of $s$. As in the model of Cui *et al.* (2000), $a$ is a shape parameter close to unity, aimed at better predicting the thermal behaviour at high temperatures. $T^0_{CT}$ is the initial temperature when the thermal loading (heating) reaches for the first time the thermal yield surface TY.

4) Other mechanisms: following the model proposed by Cui *et al.* (2000), a plastic strain takes place when a mechanical loading reaches the thermal yield curve (TY). In addition, when the mechanical yield surface (L) is reached by a thermal loading (heating), a plastic strain also occurs. In the proposed model, these plastic strains are omitted because of lack of the experimental evidences.

Regarding the hardening law of the mechanical yield surface (L), the net mean yield stress at zero suction and initial temperature, $p^*_0$, is expressed as a function of the total macrostructural plastic strain $\varepsilon^p_{vM}$:

$$\frac{\mathrm{d}p^*_0}{p^*_0} = \frac{(1+e_M)\mathrm{d}\varepsilon^p_{vM}}{\lambda(0) - \kappa} \qquad [12]$$

where $\varepsilon^p_{vM}$ is the sum of all the macrostructural plastic strains due to thermal, hydric and mechanical loading as well as those due to micro-macrostructural interactions:

$$\mathrm{d}\varepsilon^p_{vM} = \mathrm{d}\varepsilon^p_{vMT} + \mathrm{d}\varepsilon^p_{vML} + \mathrm{d}\varepsilon^p_{vMSI} + \mathrm{d}\varepsilon^p_{vMSD} \qquad [13]$$

## Model validation

Existing experimental data on compacted MX80 clay were used for the determination of parameters and model validation. Delage *et al.* (2006) investigated the soil microstructure using mercury intrusion porosimetry (MIP). Their results were used to determine the initial micro- and macrostructural void ratios ($e_m$ and $e_M$). Montes-H *et al.* (2003) observed the swelling of clay aggregate under relative humidity changes using the technique of environmental scanning electron microscopy (ESEM) in conjunction with a digital image analysis. The observed volume changes were used to calibrate the variation of $e_m$ with suction changes. The isotropic cell developed by Tang *et al.* (2007) was used to study the thermo-mechanical behaviour of unsaturated (Tang et al., 2008) and saturated (Tang & Cui, 2006) MX80 clay. The data obtained were used for the determination of parameters and model validation.

*Determination of parameters*

Fig. 2 presents the pore size distribution of a compacted MX80 clay sample having an initial suction of 113 MPa and a void ratio of 0.650 (after Delage et al., 2006). The incremental pore volume was calculated as the change of void ratio $e$ divided by the change of logarithm of pore radius $r$: $de/d\log r$, representing the total volume of the pores having similar entrance pore radius $r$. A marked double structure can be observed delimited by a pore radius of about 0.2 μm. Note that Lloret et al. (2003) obtained a similar value on compacted Febex bentonite. Thus, the micro- and macrostructural void ratios can be determined using the pore radius of 0.2 μm as the limit between macropores (inter-aggregates) and micropores (intra-aggregates): $e_m = 0.430$ and $e_M = 0.220$, as indicated in Fig. 2.

In the laboratory tests presented by Tang et al. (2008), the initial suction $s_0$ was 110 MPa and the corresponding void ratio was 0.519. Delage et al. (2006) and Lloret et al. (2003) tested soils that had a similar suction and a similar microstructural void ratio. In the present work, the microstrucutural void ratio established by Delage et al. (2006) was used, $e_m = 0.430$, and the macrostructural void ratio was deduced from $e_m$ and the total void ratio ($e = 0.519$): $e_M = 0.089$.

Montes-H et al. (2003) studied the swelling/shrinkage of MX80 aggregates at different relative humidity states. The data were used to calibrate the parameters $\alpha_m$ and $\beta_m$ which define the microstructural bulk modulus $K_m$ (Eq. [3]). In the work of Montes-H et al. (2003), an initial relative humidity of 2.5% and an initial temperature of 50°C were considered, which correspond to an initial suction of 537 MPa (see Tang & Cui, 2005). After equilibrium, various relative humidity states were applied at a temperature of 9°C, and the degree of swelling was calculated from the digital image analysis. The experimental data are presented in Fig. 3 where the volumetric strain of the aggregates (microstructural volumetric strain) is plotted versus suction calculated from the corresponding relative humidity and temperature applied. The results show that wetting from the initial suction of 537 MPa to a suction of 13.5 MPa (90% relative humidity) induced a microstructural volumetric swelling of 33%.

The following parameters were used to fit the microstructural volumetric strain to the experimental data following Eq. [3]: $\alpha_m = 0.015$ MPa$^{-1}$ and $\beta_m = 0.005$ MPa$^{-1}$. The initial suction was $s_0 = 110$ MPa and the initial microstructural void ratio was $e_m = 0.430$. The wetting and drying paths were simulated: $e_m$ reduced to 0.336 at $s = 537$ MPa and $e_m$ increased to 0.860 at $s = 0.1$ MPa. With the parameters chosen, the model satisfactorily fitted the experimental data (Fig. 3).

The parameters used for simulating the experimental results presented by Tang & Cui (2006) for a saturated state and Tang et al. (2008) for an unsaturated state are summarized in Table 1. In total, 12 simulations were performed and their stress paths are presented in Table 2. All the simulations start with the same initial conditions: $p_{ini} = 0.1$ MPa, $s_0 = 110$ MPa, and $T_0 = 25$°C.

*Swelling upon wetting*

A simulation of the volume change behaviour when wetting under zero stress was performed. From the initial state ($p_{ini}$ = 0.1 MPa, $s_0$ = 110 MPa, and $T_0$ = 25°C), $s$ was decreased to 0.1 MPa. The simulation results are plotted in Fig. 4. The void ratio values corresponding to the suctions values of 110, 39 and 9 MPa reported by Tang et al. (2008) are also presented. In the work carried out by Tang & Cui (2006), a compacted MX80 sample having similar initial state was saturated by injecting distilled water. A test was conducted in the isotropic cell described by Tang et al. (2007) under low stress, $p$ = 0.1 MPa. The final void ratio ($e$ = 1.712) obtained when the swelling was stabilized is also shown in the same figure. In this representation, the zero suction is taken corresponding to a position at 0.1 MPa.

It can be noted that the model can satisfactorily reproduce the swelling behaviour under low stress ($p$ = 0.1 MPa), even though the microstructural behaviour was fitted to the data from Montes-H et al. (2003) (Fig. 3). During the wetting path, the macrostructural behaviour involves two mechanisms: (i) elastic volumetric strain (Eq. [4]) which depends on the parameter $\kappa_s$; (ii) plastic volumetric strain induced by the micro- macrostructural interaction (Eq. [10]) which depends on $f_D$.

### *Volume change upon mechanical loading at constant suction and at a temperature of 25°C*

In order to model the effect of suction on the compressibility at a temperature of 25°C, four compression tests corresponding to four suction values were simulated. The stress paths of these simulations are presented in Fig. 5. From the initial state ($p_{ini}$ = 0.1 MPa, $s_0$ = 110 MPa, and $T_0$ = 25°C), $p$ was increased to 60 MPa for the simulation S01. For the other simulations, $s$ was first decreased to 39 (S02), 9 (S03) and 0.1 MPa (S04) while $p$ was kept constant at 0.1 MPa. After this wetting path, $p$ was increased to 10 MPa (for S02) and to 5 MPa (for S03 and S04).

The results of these simulations are presented in Fig. 6, together with the experimental data. For all four suctions ranging from 0.1 to 110 MPa, the model is in good agreement with the experimental data. Note that during the mechanical loading, the following volumetric strains were taken into account: $\varepsilon_{vm}^e$ (Eq. [2]), $\varepsilon_{vM}^e$ (Eq. [4]), $\varepsilon_{vML}^p$ (Eq. [8]) and $\varepsilon_{vMSI}^p$ (Eq. [9]). With the exception of simulation S01 at high pressure, the other simulations showed negligible microstructural volumetric strain upon loading.

### *Volume change upon thermal loading*

The stress paths considered for studying the thermal volumetric behaviour are presented in Fig. 7. From the initial state, $T$ was increased to 80°C for S05. For S06, $s$ was first decreased to 39 MPa and then $T$ was increased to 80°C. For S07, wetting to $s$ = 9 MPa was followed by heating to 80°C. For S08, $p$ was increased first to 5 MPa prior to heating. For S09, after wetting to $s$ = 39 MPa, $p$ was increased to 5 MPa before heating to $T$ = 80°C.

In Fig. 8, the volumetric strains induced by heating are presented for all these simulations together with the corresponding experimental data. Note that in these simulations, the coefficients of thermal expansion were taken negative (see Table 1). The main responses observed experimentally by Tang et al. (2008) were reproduced satisfactorily. For instance, heating at high suction ($s$ = 110 MPa, S05 and S08) induced only a linear expansion; however heating at lower suction ($s$ = 39 and 9 MPa) induced a thermal expansion followed by a thermal contraction. For a same suction ($s$ = 39 MPa), the temperature limit, where the thermal loading yield surface *TY* was reached, was higher at a lower stress ($T_{CT}^0$ = 58°C with *p* = 0.1 MPa, S06; $T_{CT}^0$ = 29°C with *p* = 5 MPa, S09). When comparing S06 and S07 having the same net mean stress, *p* = 0.1 MPa, $T_{CT}^0$ was lower at a lower suction ($T_{CT}^0$ = 25°C for *s* = 9 MPa, S07).

*Temperature and suction effect on the compressibility*

Fig. 9 presents the stress paths of the simulations that aim at studying the effect of suction and temperature on the compressibility. In addition to the three simulations at a temperature of 25°C presented previously (S01, S02, and S03), three other mechanical loading paths were considered at higher temperatures. For S10, heating to 60°C was followed by a mechanical loading to *p* = 60 MPa. An initial wetting to *s* = 39 MPa was first applied for S11 before heating to *T* = 60°C. Finally, loading to *p* = 10 MPa was performed. For S12, after wetting to *s* = 9 MPa, a thermal cycle was applied: *T* = 25 – 80 – 25 – 80°C. Finally, at *T* = 80°C, a mechanical loading was undertaken to *p* = 5 MPa. Following the experimental results (Tang *et al.*, 2008), heating from 25 to 80°C (at *s* = 9 MPa, *p* = 0.1 MPa) induced a plastic compression and the subsequent thermal cycle 80 – 25 – 80°C induced a reversible behaviour. In the proposed model, after the heating from 25 to 80°C, the subsequent cooling phase (from 80 to 25°C) induced elastic thermal volumetric strains (equations [2] and [4]) but these elastic volumetric strains did not induce any plastic strain following equations [9] and [10]. As a result, the thermal cycle 80 – 25 – 80°C equally induced a reversible behaviour in the simulation. Therefore, the difference in void ratio between simulation S03 and S12 (Fig. 10*c*) at *p* = 0.1 MPa corresponds to the thermal compression of 1.3% obtained in the simulation S07 (Fig. 8*c*).

The compression curves (*e* versus *p*) at constant *s* and *T* are presented in Fig. 10 for all these simulations together with the corresponding experimental data. Except for the curves at *s* = 39 MPa (S02 and S11), the model is in satisfactory agreement with the experimental data; the main responses observed experimentally were well described: $p_0$ is lower at a lower suction or a higher temperature; the compression slopes are independent of the temperature.

*Evolution of the yield surfaces*
For further understanding the mechanisms of the proposed model, the evolution of the yield surfaces for simulation S12 is plotted in Fig. 11. The loading surface is represented by the *s-p* plot for T = 25°C (Fig. 11*b*) and by the *T-p* plot for *s* = 0.1 MPa (Fig. 11*d*). The thermal surface is represented by the *T-s* plot for *p* = 0.1 MPa (Fig. 11*a*) and by the *T-p* plot for *s* = 0.1 MPa (Fig. 11*c*). The positions of the thermal and loading yield surfaces were plotted for four states: (i) initial state, *s* = 110 MPa, *T* = 25°C, *p* = 0.1 MPa; (ii) end of the wetting path, *s* = 9 MPa, *T* = 25°C, *p* = 0.1 MPa; (iii) end of the heating path, *s* = 9 MPa, *T* = 80°C, *p* = 0.1 MPa; (iv) end of the loading path, *s* = 9 MPa, *T* = 80°C, *p* = 5 MPa. It can be observed that from the initial state, wetting (suction decreased from 110 to 9 MPa) induced a decrease of $p_0^*$ from 0.50 to 0.20 MPa, leading to a movement of the loading yield surface to the left (Fig. 11*b,d*) following the equation [5]. As the thermal yield surface is mainly governed by $p_0$, according to equation [7], a decrease of $p_0$ moved it to the right in Fig. 11*a* and to the left in Fig. 11*c*. The subsequent heating (T = 25 – 80°C) induced plastic volumetric strains, and as a result, increased $p_0^*$ from 0.20 to 0.25 MPa. That moved the loading yield surface to the right in Fig. 11*b,d* and the thermal yield surface to the left in Fig. 11*a* and to the right in Fig. 11*c*. Finally, when the loading (*p* = 0.1 – 5 MPa) was performed at *s* = 9 MPa, *T* = 80°C, plastic volumetric strains occurred. That increased $p_0^*$ from 0.25 MPa to 1.00 MPa, further moving the loading yield surface to the right in Fig. 11*b,d* and the thermal yield surface to the left in Fig. 11*a* and to the right in Fig. 11*c*.

## Conclusions

A constitutive model was elaborated to describe the thermo-mechanical volume change behaviour of compacted expansive clays. This model is based on the model proposed by Alonso *et al.* (1999) for the hydro-mechanical behaviour of compacted unsaturated expansive clays and the model proposed by Cui *et al.* (2000) for the thermo-mechanical behaviour of saturated clays. Existing experimental data obtained by Montes-H *et al.* (2003), Delage *et al.* (2006) and Tang *et al.* (2008) on compacted MX80 clay were used for the determination of parameters and model validation. The simulations performed show that the main volume change behaviour observed experimentally was satisfactorily reproduced: swelling upon wetting at low stress ($p$ = 0.1 MPa) and at ambient temperature; compressibility under various suctions and temperatures; elastic thermal expansion and plastic thermal contraction under various suctions and stresses.

## Notation

$a$      parameter controlling the plastic volumetric strain at macrostructural level when the thermal yield surface is reached by thermal loading
$e$      void ratio
$e_m$      microstructural level void ratio
$e_M$      macrostrucutural level void ratio
$f_I$      interaction function between micro- and macrostructural levels for suction increasing
$f_D$      interaction function between micro- and macrostructural levels for suction decreasing
$K_m$      microstructural bulk modulus for changes in suction plus mean stress
$OCR$      over-consolidation ratio
$p$      net mean stress
$p_{atm}$      atmospheric pressure (0.1 MPa)
$p_c$      reference stress
$p_0$      net mean yield stress at current suction and temperature
$p_O^*$      net mean yield stress at zero suction and initial temperature ($T_0$)
$p_{ini}$      initial net mean stress
$r$      parameter defining the maximum macrostructural soil compressibility
$s$      suction
$s_0$      initial suction
$T$      temperature
$T_0$      initial temperature
$T_c$      reference temperature
$T_{CT}$      yield temperature at current pressure and suction
$T_{CT}^0$      initial yield temperature when the thermal loading (heating) reaches the thermal yield surface TY for the first time
$\alpha_0$ $\alpha_0$      parameter controlling the decreasing rate of net mean yield stress with temperature

$\alpha_m$   parameter controlling the increasing rate of microstructural soil stiffness with mean stress plus suction

$\alpha_M^T$   coefficient of thermal expansion at macrostructural level

$\alpha_m^T$   coefficient of thermal expansion at microstructural level

$\alpha_s$   parameter controlling the decreasing rate of yield temperature with decrease of suction

$\beta$   parameter controlling the increasing rate of macrostructural soil compressibility with suction

$\beta_m$   parameter controlling the microstructural soil stiffness

$\beta_T$   parameter controlling the decreasing rate of yield temperature with net mean stress

$\varepsilon_v$   total volumetric strain

$\varepsilon_{vm}^e$   elastic volumetric strain at microstructural level

$\varepsilon_{vM}^e$   elastic volumetric strain at macrostructural level

$\varepsilon_{vML}^p$   plastic volumetric strain at macrostructural level when the yield surface L is reached by mechanical loading

$\varepsilon_{vMSI}^p$   plastic volumetric strain at macrostructural level induced by the micro-macrostructural interaction with suction increase

$\varepsilon_{vMSD}^p$   plastic volumetric strain at macrostructural level induced by the micro-macrostructural interaction with suction decrease

$\varepsilon_{vMT}^p$   plastic volumetric strain at macrostructural level when the thermal yield surface is reached by thermal loading

$\kappa$   macrostructural elastic compressibility parameter for changes in net mean stress

$\kappa_s$   macrostructural elastic compressibility parameter for changes in suction

$\lambda(0)$   macrostructural compressibility parameter for changes in net mean stress for virgin states of soil at zero suction

$\lambda(s)$   macrostructural compressibility parameter for changes in net mean stress for virgin states of soil at suction $s$

**Table 1. Parameters used in modelling.**

| Initial state | | Hydro-mechanical behaviour | | Thermal behaviour | |
|---|---|---|---|---|---|
| $e$ | 0.519 | $\kappa$ | 0.008 | $\alpha_m^T$ | - 0.0001°C$^{-1}$ |
| $e_M$ | 0.430 | $\kappa_s$ | 0.077 | $\alpha_M^T$ | - 0.0001°C$^{-1}$ |
| $e_m$ | 0.089 | $\lambda(0)$ | 0.25 | $\alpha_0$ | 0.01°C$^{-1}$ |
| $T_0$ | 25°C | $p_c$ | 0.25 MPa | $T_c$ | 200°C |
| $s_0$ | 110 MPa | $r$ | 0.18 | $\alpha_s$ | 0.016 MPa$^{-1}$ |
| $p_{ini}$ | 0.1 MPa | $\beta$ | 0.14 | $\beta_T$ | 0.9 |
| $p_0^*$ | 0.5 MPa | $\alpha_m$ | 0.015 MPa$^{-1}$ | $a$ | 0.975 |
| | | $\beta_m$ | 0.005 MPa$^{-1}$ | | |
| | | $f_I = 0.05 + 0.1\tanh[20(p/p_0) - 0.25]$ | | | |
| | | $f_D = 0.75 + 0.05\tanh[20(p/p_0) - 0.25]$ | | | |

**Table 2. Stress paths for simulations**

| Simulation No. | Path* | | | | | | | | | | | | | | |
|---|---|---|---|---|---|---|---|---|---|---|---|---|---|---|---|
| | I | | | II | | | III | | | IV | | | V | | |
| | *p*: MPa | *s*: MPa | *T*: °C | *p*: MPa | *s*: MPa | *T*: °C | *p*: MPa | *s*: MPa | *T*: °C | *p*: MPa | *s*: MPa | *T*: °C | *p*: MPa | *s*: MPa | *T*: °C |
| S01 | 60 | 110 | 25 | | | | | | | | | | | | |
| S02 | 0.1 | 39 | 25 | 10 | 39 | 25 | | | | | | | | | |
| S03 | 0.1 | 9 | 25 | 5 | 9 | 25 | | | | | | | | | |
| S04 | 0.1 | 0.1 | 25 | 5 | 0.1 | 25 | | | | | | | | | |
| S05 | 0.1 | 110 | 80 | | | | | | | | | | | | |
| S06 | 0.1 | 39 | 25 | 0.1 | 39 | 80 | | | | | | | | | |
| S07 | 0.1 | 9 | 25 | 0.1 | 9 | 80 | | | | | | | | | |
| S08 | 5 | 110 | 25 | 5 | 110 | 80 | | | | | | | | | |
| S09 | 0.1 | 39 | 25 | 5 | 39 | 25 | 5 | 39 | 80 | | | | | | |
| S10 | 0.1 | 110 | 60 | 60 | 110 | 60 | | | | | | | | | |
| S11 | 0.1 | 39 | 25 | 0.1 | 39 | 60 | 10 | 39 | 60 | | | | | | |
| S12 | 0.1 | 9 | 25 | 0.1 | 9 | 80 | 0.1 | 9 | 25 | 0.1 | 9 | 80 | 5 | 9 | 80 |

* Stress paths start at a common initial state ($p = 0.1$ MPa, $s = 110$ MPa, $T = 25$°C). The Table indicates the subsequent ($p, s, T$) values; each path corresponds to a loading at constant *s* and *T*, or a *T* change at constant *s* and *p*.

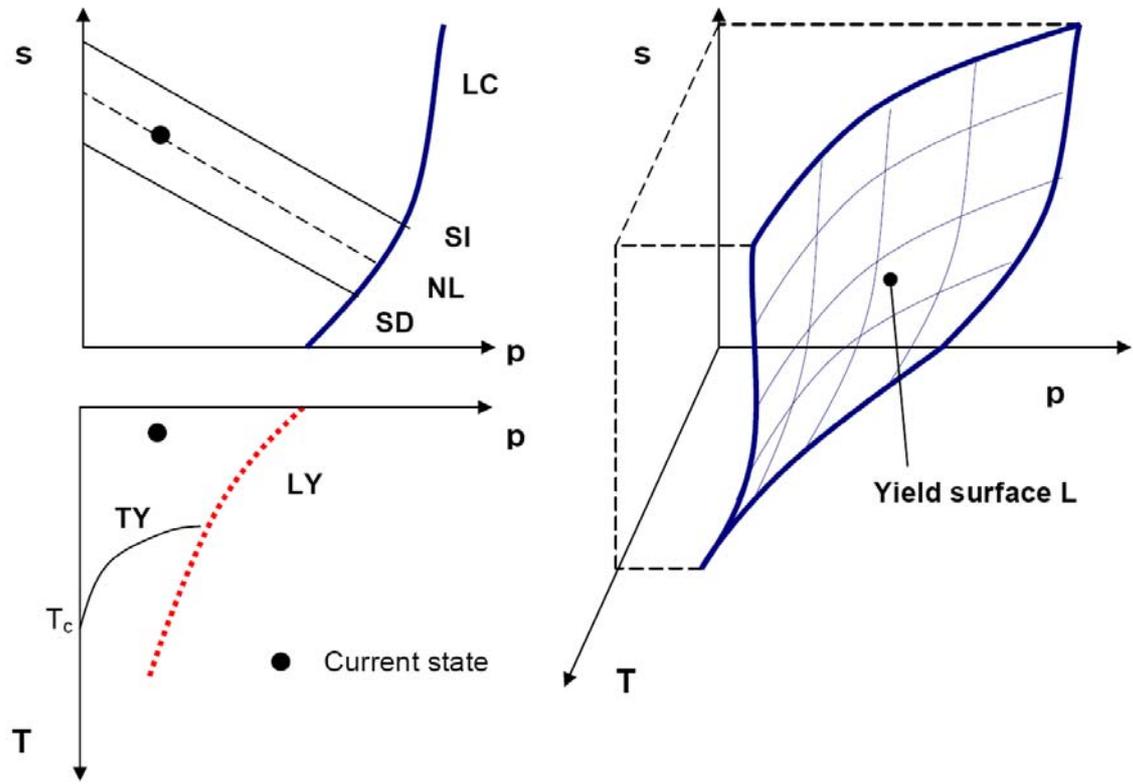

**Fig. 1. Schematic view of the yield surfaces**

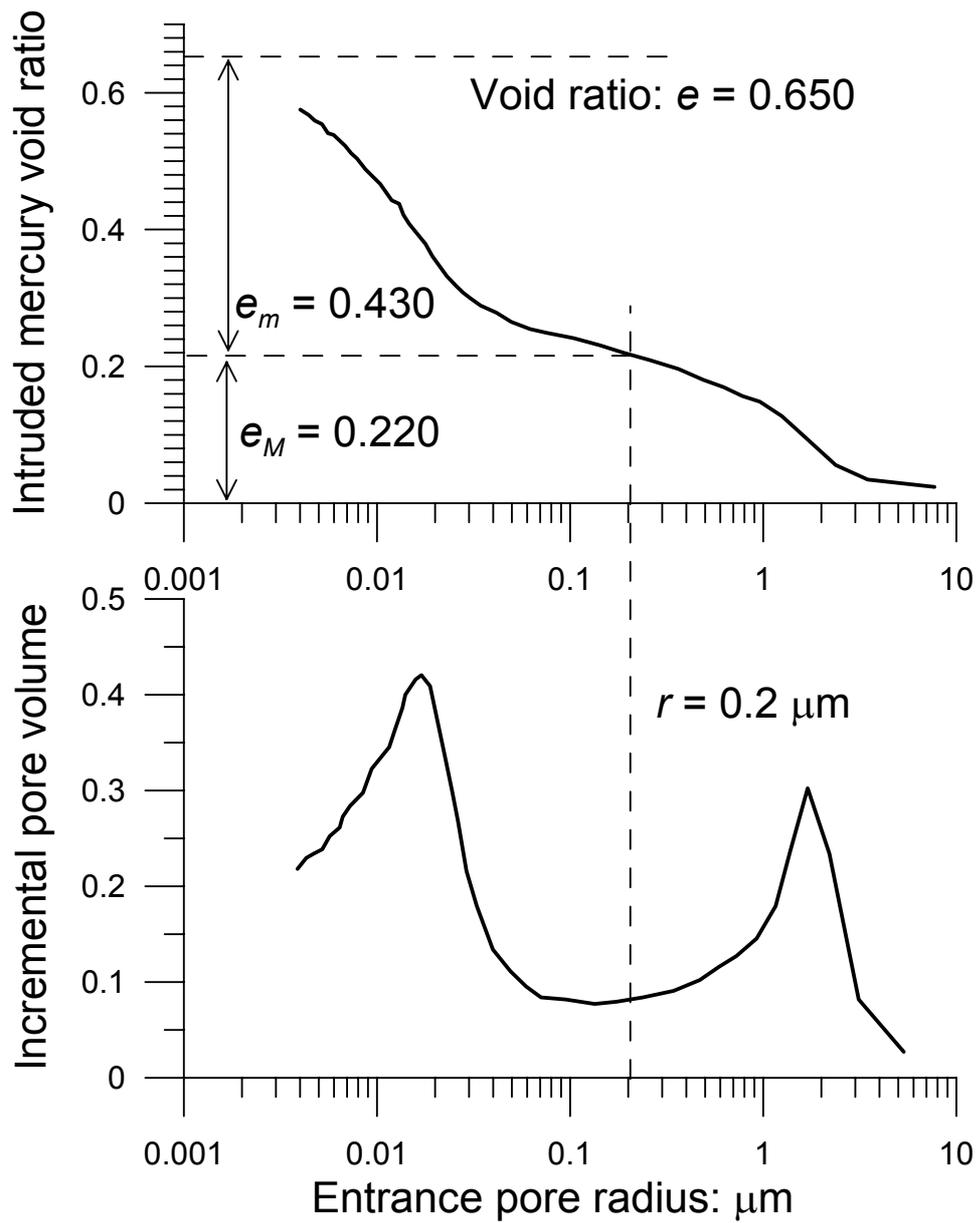

**Fig. 2. Pore size distribution of compacted MX80 clay (after Delage *et al.*, 2006)**

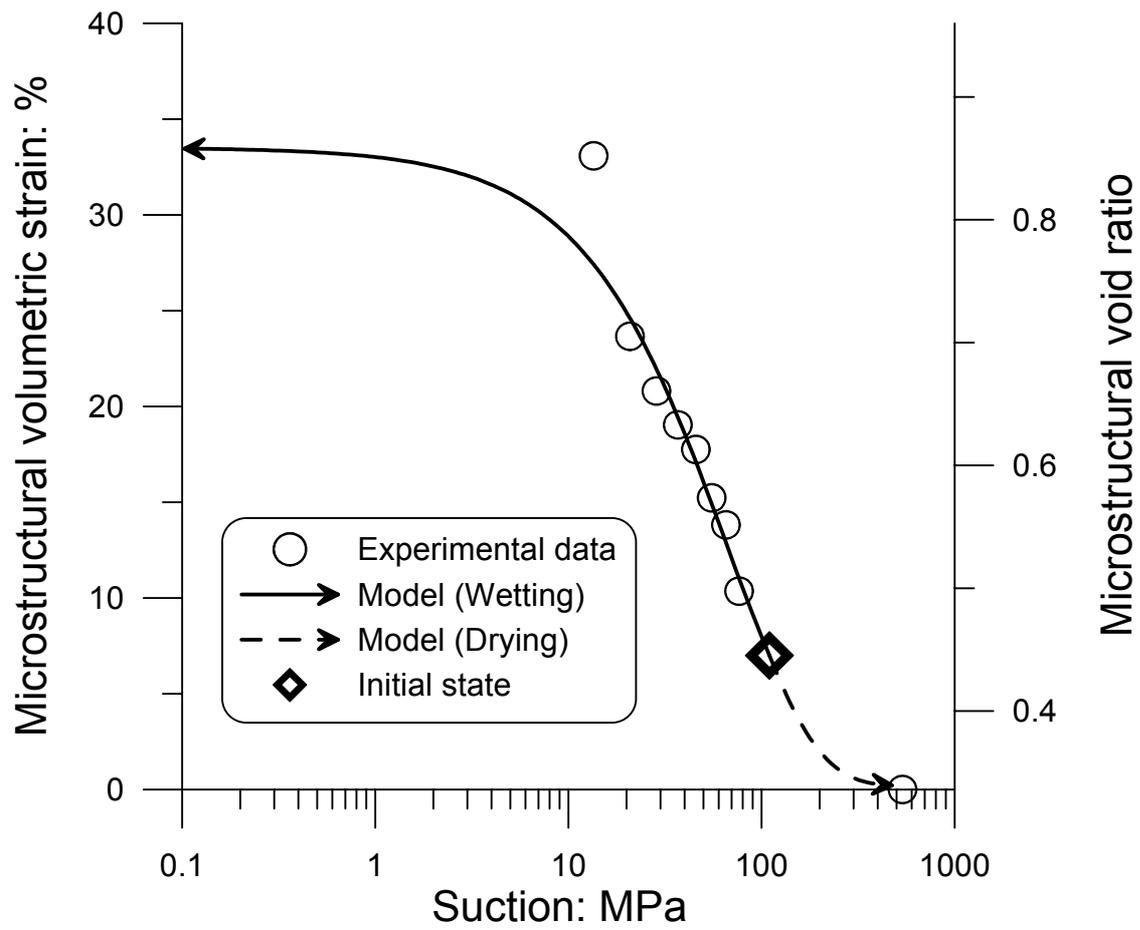

**Fig. 3. Microstructural volumetric strain and microstructural void ratio versus suction**

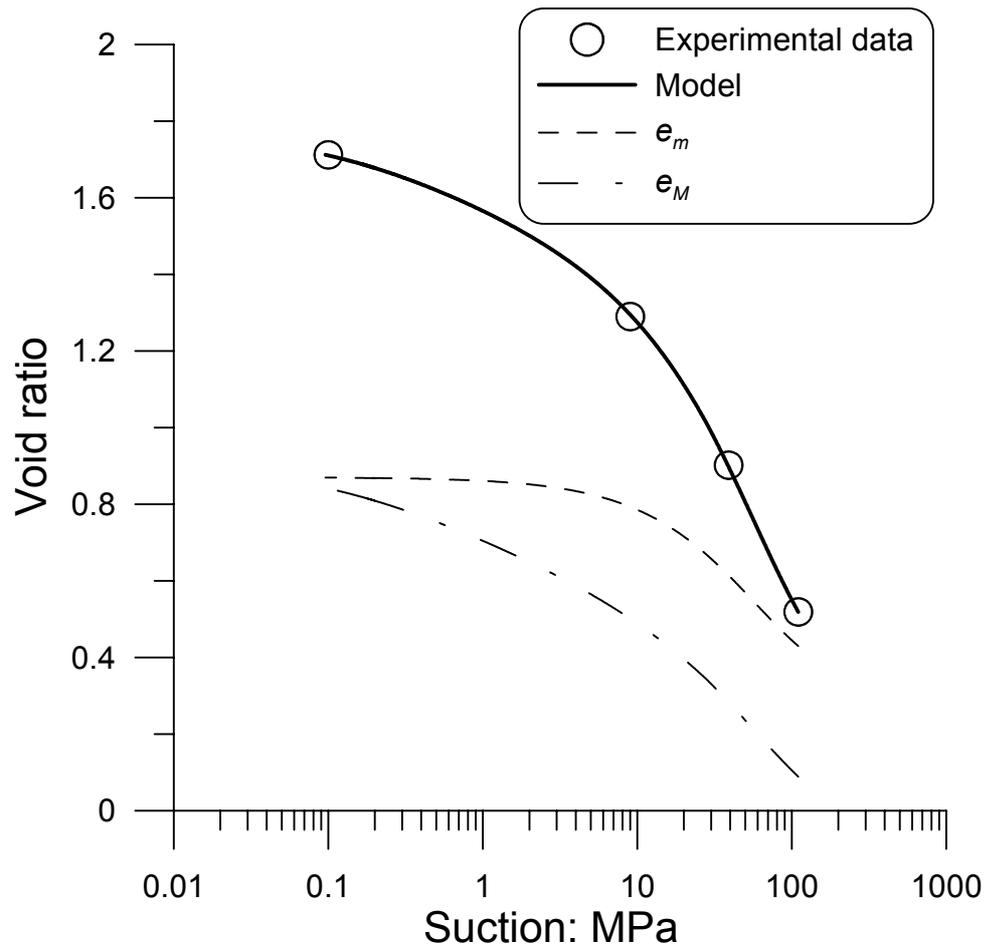

**Fig. 4. Void ratio changes during wetting at zero stress**

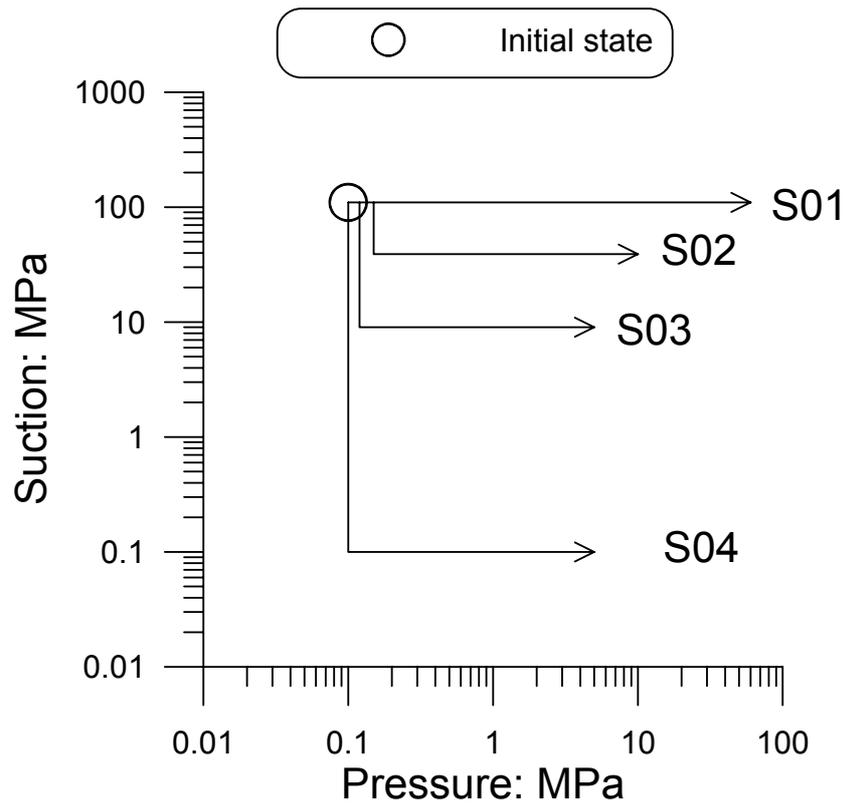

**Fig. 5.** Stress paths of simulations S01, S02, S03, and S04 (for studying the behaviour under mechanical loading at constant suction and at a temperature of 25°C)

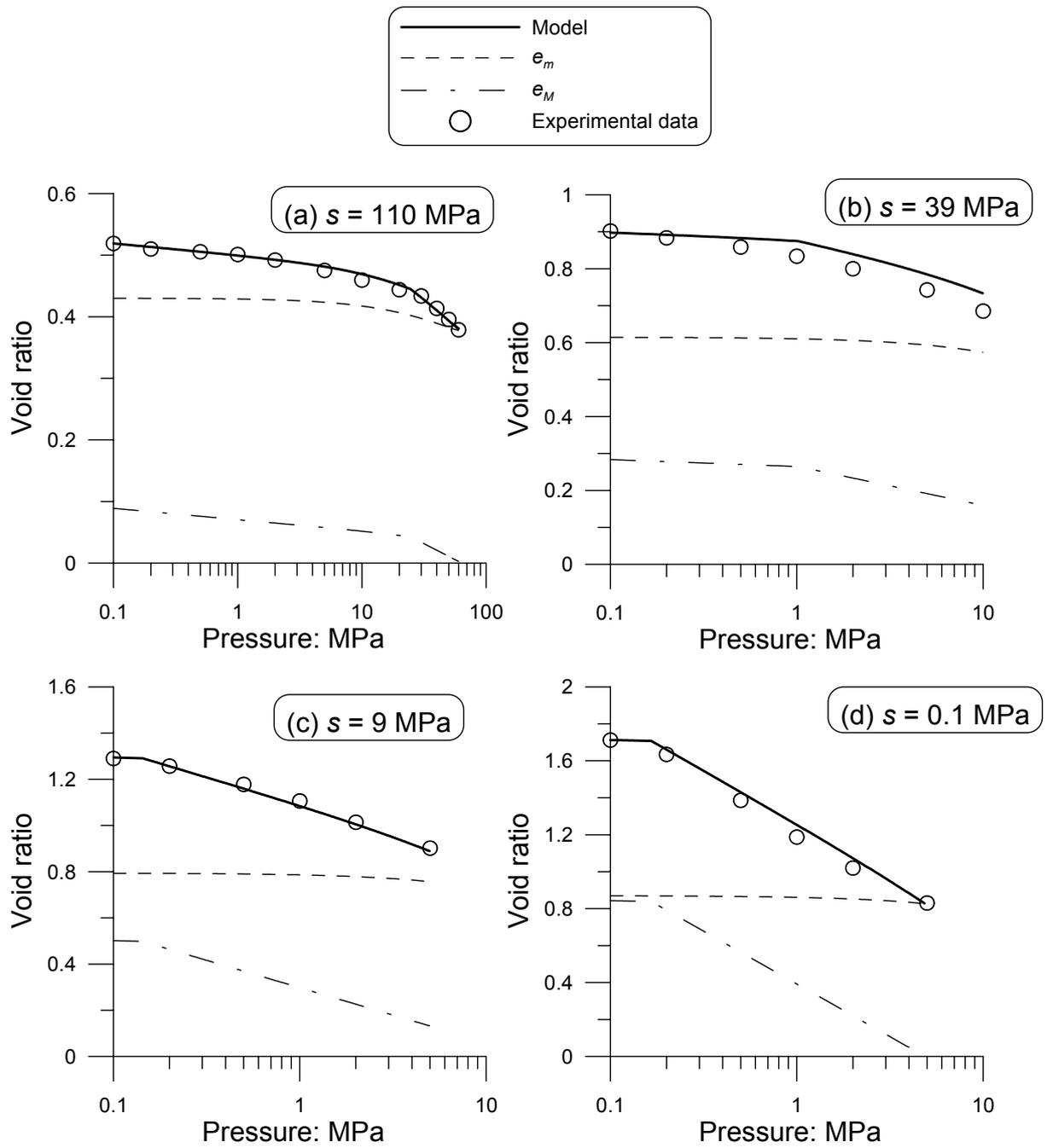

**Fig. 6. Void ratio changes during mechanical loading at constant suction and at a temperature of 25°C**

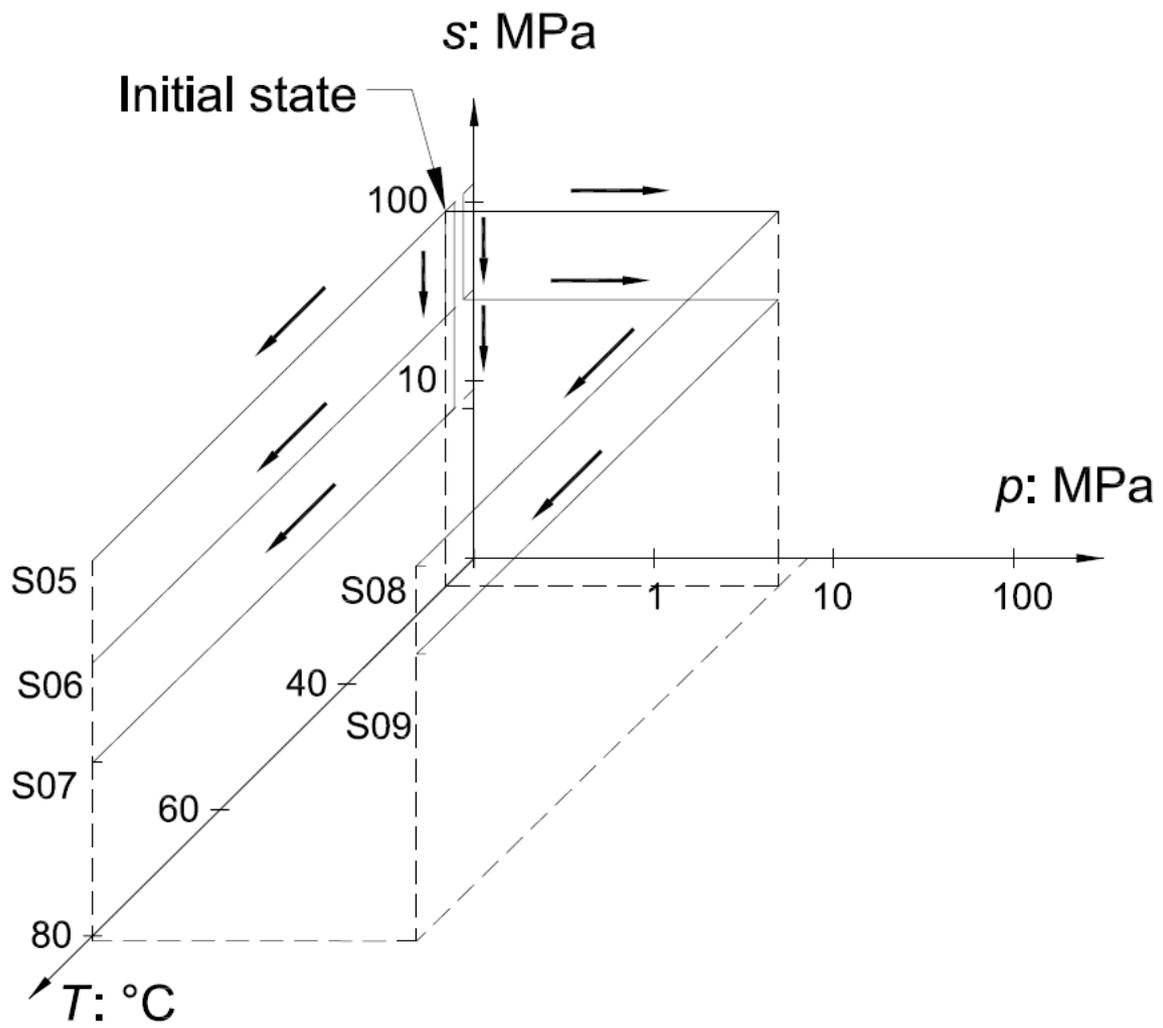

Fig. 7. Stress paths of simulations S05, S06, S07, S08, and S09 (for studying the thermal volumetric behaviour)

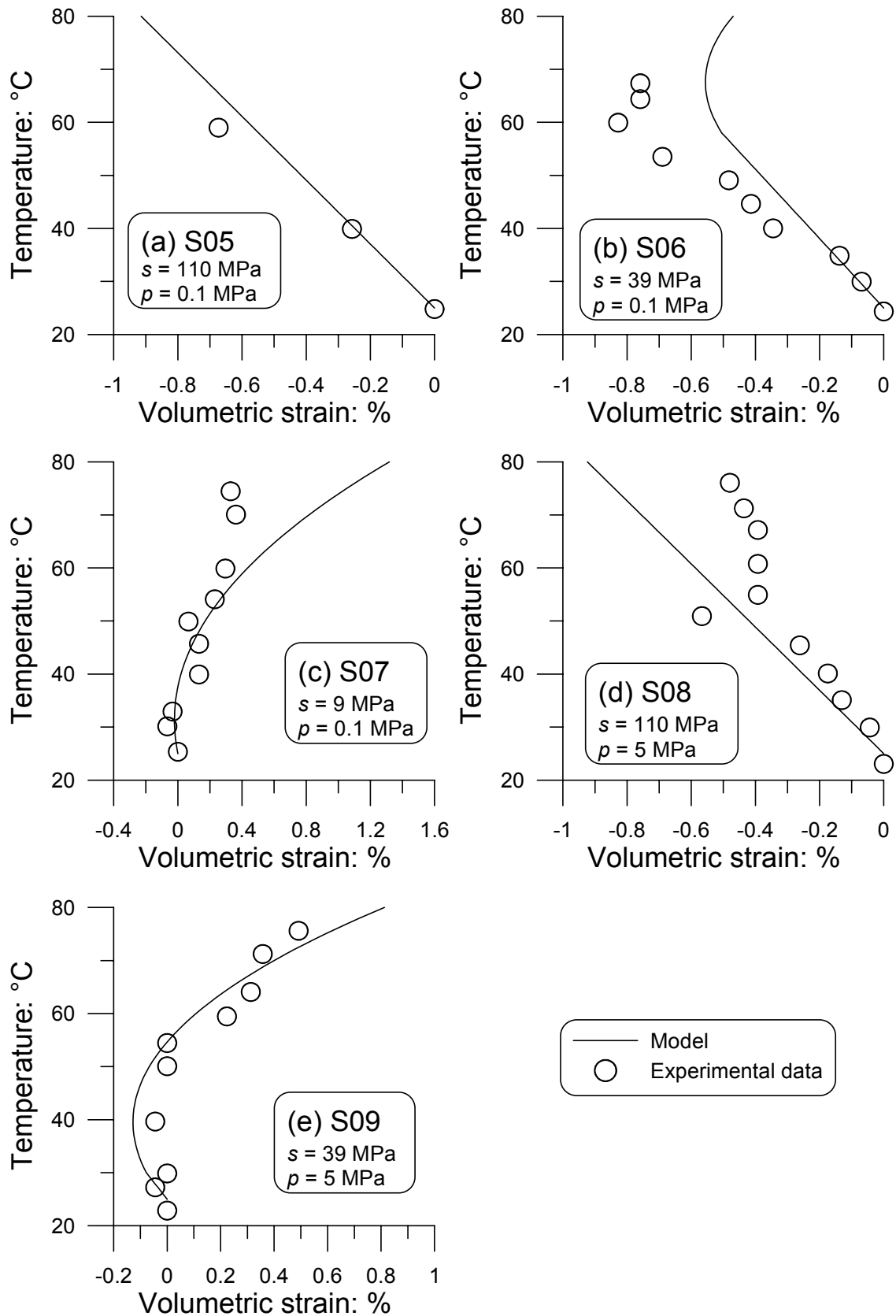

**Fig. 8.** Volumetric strain during heating (from 25 to 80°C) under constant pressures for various suction values: simulations S05, S06, S07, S08 and S09

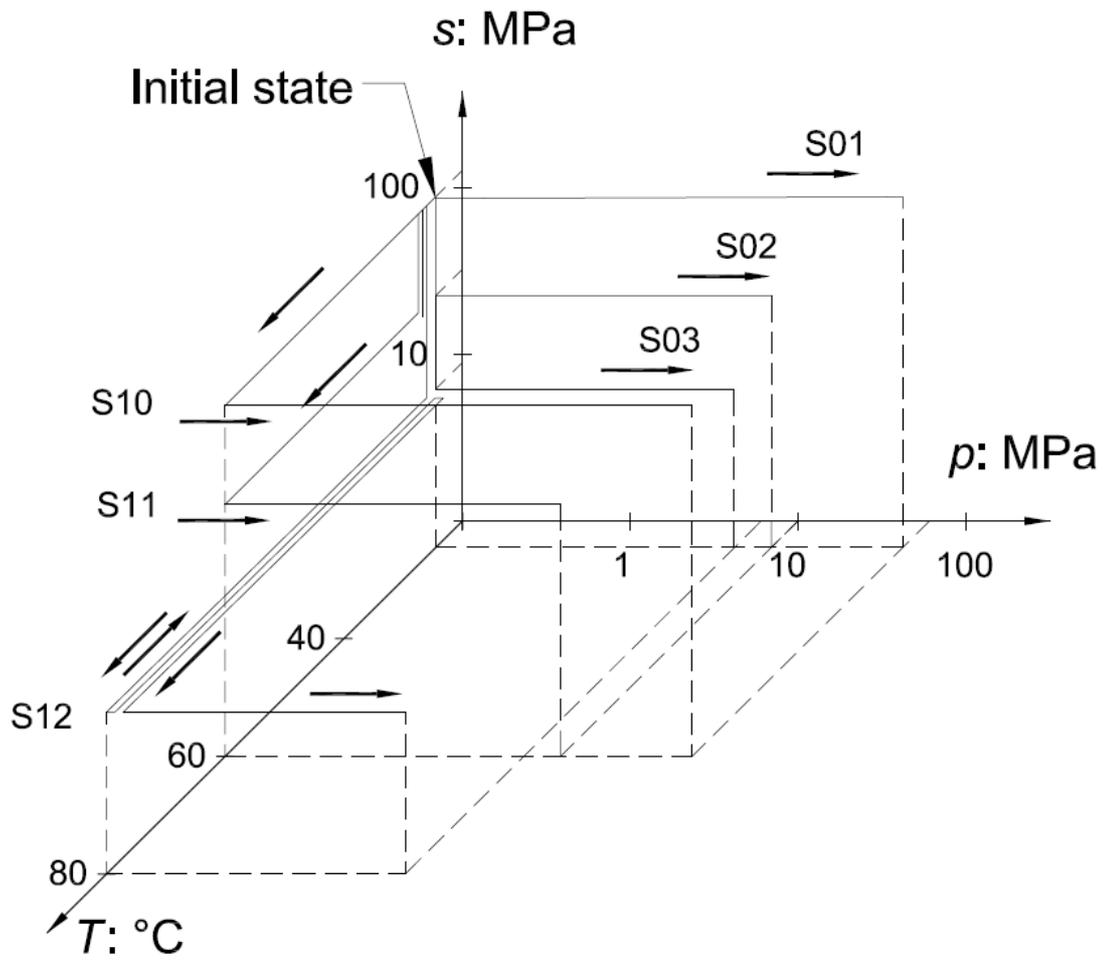

**Fig. 9.** Stress paths for simulations S01, S02, S03, S10, S11, and S12 (for studying mechanical volumetric behaviour)

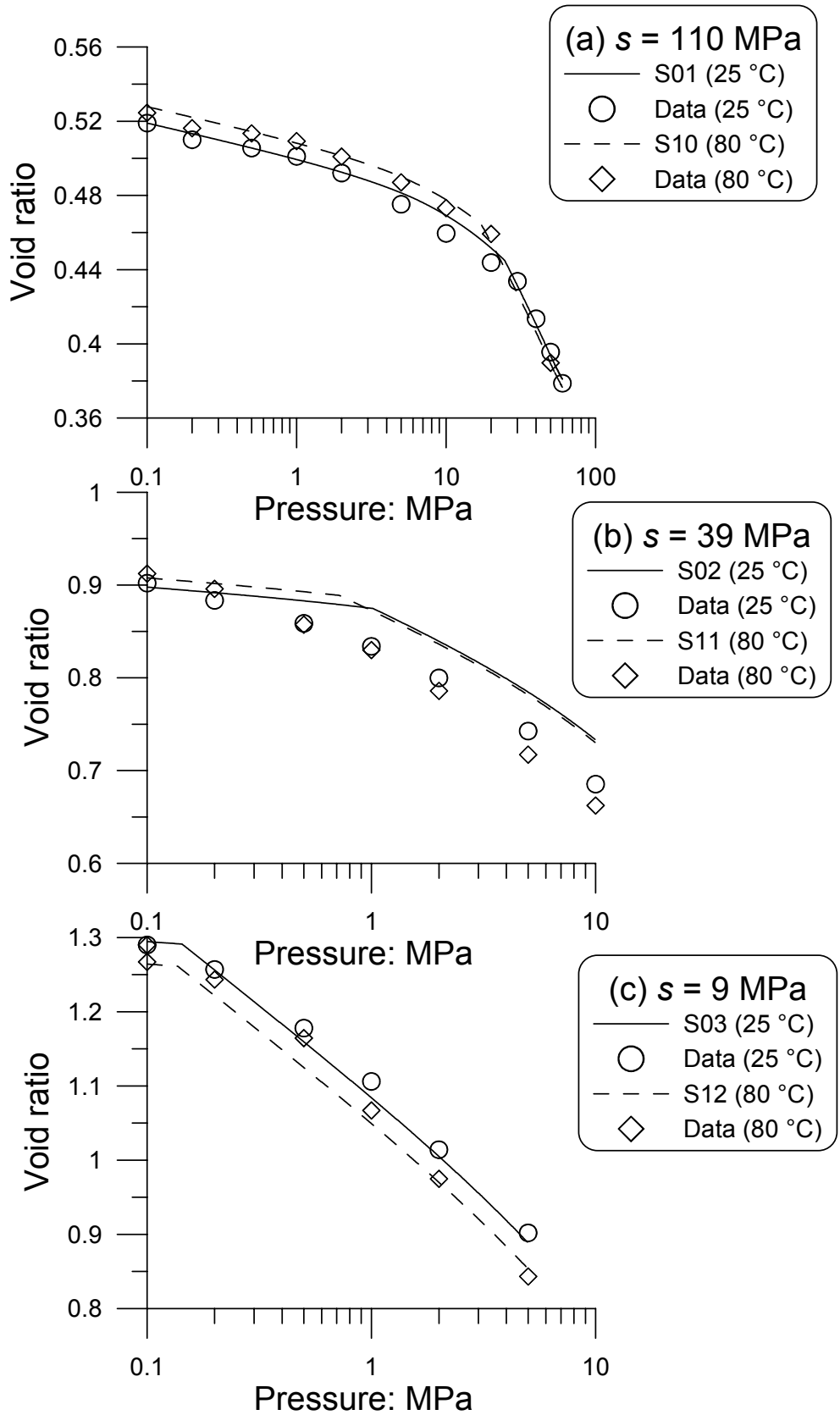

**Fig. 10. Void ratio changes during mechanical loading at constant suction and temperature: simulations S01, S02, S03, S10, S11, and S12**

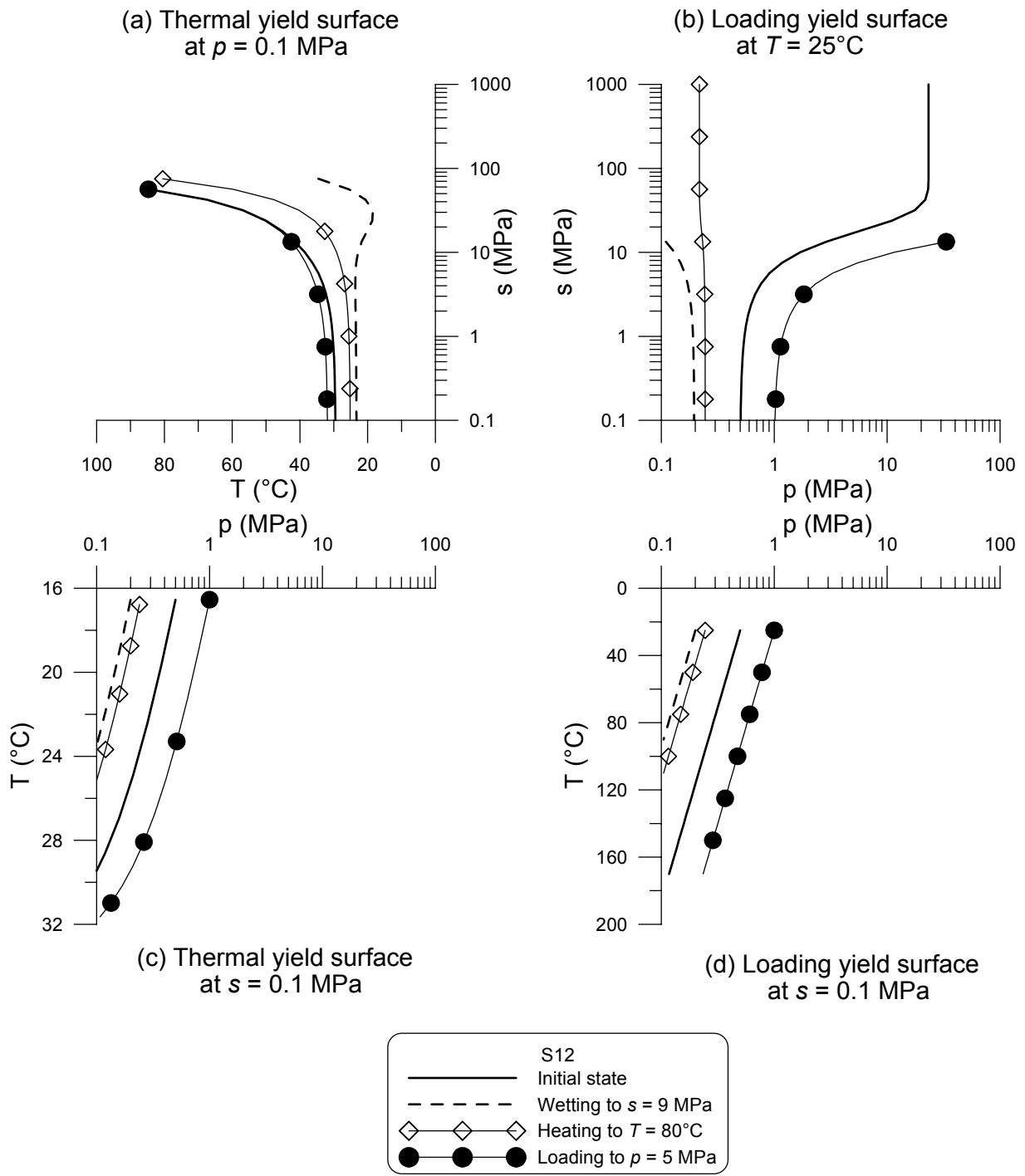

Fig. 11. Evolution of the yield surfaces in simulation S12